\documentclass[runningheads]{llncs}
\usepackage[T1]{fontenc}
\usepackage{graphicx}
\usepackage{cite}
\usepackage{multirow}
\usepackage{amsmath} 
\usepackage{booktabs}
\usepackage{threeparttable}
\usepackage{wrapfig}
\usepackage{amsfonts}
\usepackage{subcaption}
\usepackage{tikz}
\usepackage{circledsteps}

\begin{document}
\title{An Interpretable and Attention-based Method for Gaze Estimation Using Electroencephalography}
%
%

\author{Nina Weng\inst{1}
\and
Martyna Plomecka\inst{2}
\and
Manuel Kaufmann\inst{3}
\and 
Ard Kastrati\inst{3}
\and
Roger Wattenhofer\inst{3}
\and
Nicolas Langer\inst{2}
}

\authorrunning{Weng et al.}
%
\institute{Technical University of Denmark, Denmark\\
\email{ninwe@dtu.dk}\\
\and
University of Zurich, Switzerland 
\email{martyna.plomecka@uzh.ch,n.langer@psychologie.uzh.ch}\\
\and
ETH Zurich, Switzerland\\
\email{\{kamanuel,akastrati,wattenhofer\}@ethz.ch}
}

\maketitle              

\begin{abstract}

Eye movements can reveal valuable insights into various aspects of human mental processes, physical well-being, and actions.
Recently, several datasets have been made available that simultaneously record EEG activity and eye movements. This has triggered the development of various methods to predict gaze direction based on brain activity.
However, most of these methods lack interpretability, which limits their technology acceptance.
In this paper, we leverage a large data set of simultaneously measured Electroencephalography (EEG) and Eye tracking, proposing an interpretable model for gaze estimation from EEG data. More specifically, we present a novel attention-based deep learning framework for EEG signal analysis, which allows the network to focus on the most relevant information in the signal and discard problematic channels. 
Additionally, we provide a comprehensive evaluation of the presented framework, demonstrating its superiority over current methods in terms of accuracy and robustness.
Finally, the study presents visualizations that explain the results of the analysis and highlights the potential of attention mechanism for improving the efficiency and effectiveness of EEG data analysis in a variety of applications.

\end{abstract}
\section{Introduction}

Gaze information is a widely used behavioral measure to study attentional focus \cite{eckstein2017beyond}, cognitive control \cite{munoz2004look}, memory traces \cite{ryan2010eye} and decision making \cite{vachon2014eye}.
The most commonly used gaze estimation technique in laboratory settings is the infrared eye tracker, which detects gaze position by emitting invisible near-infrared light and then capturing the reflection from the cornea \cite{duchowski2007eye}.
While infrared eye tracker still remains the most accurate and reliable solution for the gaze estimation, these systems have several limitations, including individual differences in the contrast of the pupil and iris and the need for time-consuming setup and calibration before each scanning session \cite{carter2020best, holmqvist2012eye}.

Recently, Electroencephalogram (EEG) has been explored as an alternative method to estimate eye movements by recording electrical activity from the brain non-invasively with high temporal resolution \cite{kastrati2021eegeyenet}.
The growing body of literature has shown that Deep Learning architectures could be significantly effective for many EEG-based tasks \cite{tabar2016novel,craik2019deep}.
Nevertheless, with the advantages that Deep Learning brings, new challenges arise. Most of these models applied to electroencephalography (EEG) data tend to lack \textit{interpretability}, making it difficult to understand the underlying reasons for their predictions, which subsequently leads to a decrease in the acceptability of advanced technology in neuroscience \cite{sturm2016interpretable}.
However, a potential solution already exists, in the form of the attention mechanism \cite{vaswani2017attention}. The attention mechanism has the potential to provide a more transparent and understandable way of analyzing EEG data, enabling us to comprehend the relationships between different brain signals better and make more informed decisions based on the results. With the development and implementation of these techniques, we can look forward to a future where EEG data can be utilized more effectively and efficiently in various applications.

Attention mechanisms have recently emerged as a powerful tool for processing sequential data, including time-series data in various fields such as natural language processing, speech recognition, and computer vision \cite{vaswani2017attention, shaw2018self, devlin2018bert}. In the context of EEG signal analysis, attention mechanism has shown promising results in various applications, including sleep stage classification, seizure detection, and event-related potential analysis \cite{feng2021automatic,lee2022eeg, hu2022eeg}.
Since different electrodes record the brain activity from the different brain areas and functions, the information density from each electrode can vary for different tasks \cite{kastrati2023electrode}.

In this study, we introduce a new deep learning framework for analyzing EEG signals applying attention mechanisms. 
For the method evaluation, we used the EEGEyeNet dataset and benchmark \cite{kastrati2021eegeyenet}, which includes concurrent EEG and infrared eye-tracking recordings, with eye tracking data serving as a ground truth.
Our method incorporates attention modules to assign weights to individual electrodes based on their importance, allowing the network to prioritize relevant information in the signal. Specifically, we demonstrate the ability of our framework to accurately predict gaze position and saccade direction, achieving superior performance compared to previously benchmarked methods.
Furthermore, we provide visualizations of model's interpretability through case studies.

\section{Model}

\subsection{Motivation}

In this study, our primary goal was to build a model sensitive to different electrodes. The motivation for this goal is two-fold.
Firstly, with regards to interpreting the model, the electrodes can be considered the smallest entity as they record signals from specific regions of the brain. 
Therefore, the electrode-based explanation is a reasonable approach considering human understanding.
Second, in the context of model learning, incorporating adaptive weighting of electrodes within a neural network can potentially enhance the accuracy and reliability of gaze estimation systems. This is because electrodes are functionally connected to cognitive behaviors. Specifically, in tasks such as gaze estimation, electrodes positioned near the eyes can capture electrical signals from the orbicularis oculi muscles \cite{bulling2010eye}, thereby making the pre-frontal brain areas more crucial for precise estimation \cite{kastrati2023electrode}.
Additionally, the noise of EEG recordings could be induced by 
\begin{wrapfigure}{r}{6cm}
\centering
\includegraphics[width=5.5cm]{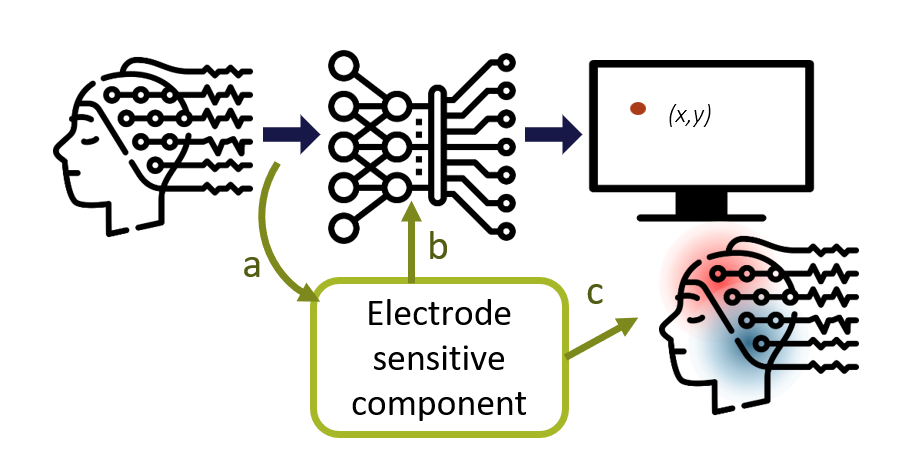}
\caption{We augment an electrode-sensitive component to a deep learning model, which works as follows: a) extract electrode-wise information from input data, b) control the predictions, and c) provide explanations.}\label{fig:framework}
\end{wrapfigure} 
broken wire contacts, too much or dried gel, or loose electrodes \cite{teplan2002fundamentals}, the influence of such electrodes should be reduced in the network under ideal circumstances.

As shown in Figure \ref{fig:framework}, our model design focuses on enhancing an existing deep learning architecture with an electrode-sensitive component. This component first extracts electrode-related information, and then utilizes this information for two purposes: (1) emphasizing the reliable electrodes and diminishing the influence of suspicious electrodes, while simultaneously (2) providing explanations for each prediction.

\subsection{Attention-CNN}

\begin{figure}[b]
\centering
\includegraphics[width=0.98\textwidth]{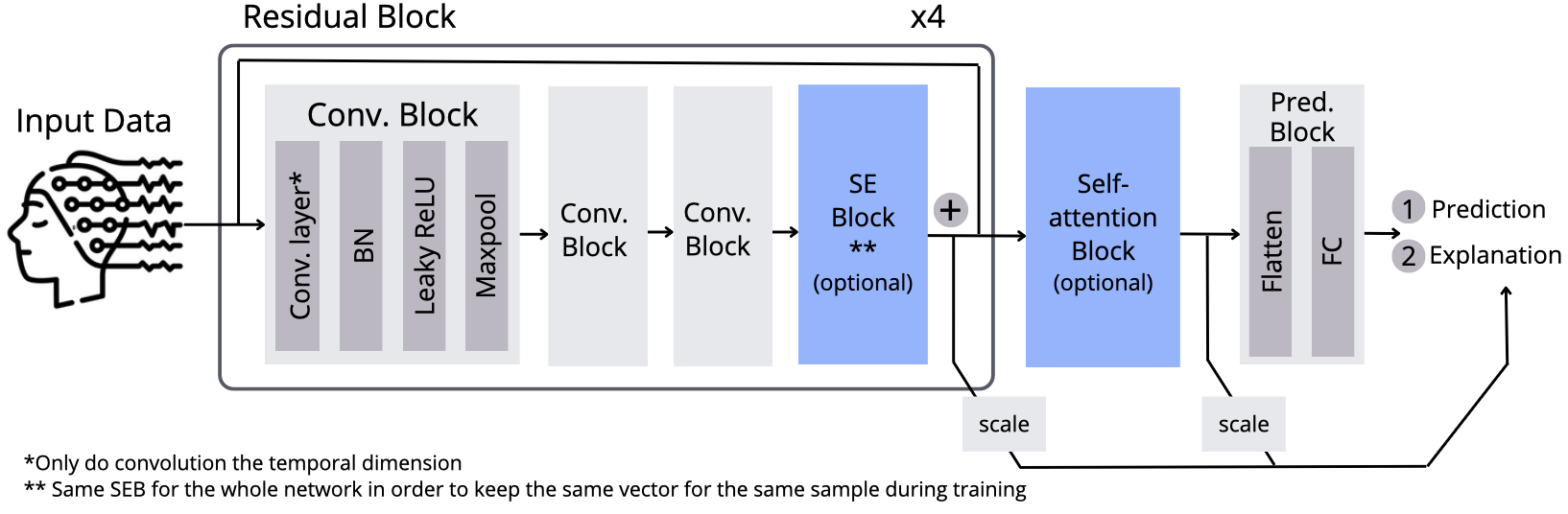}
\caption{The Architecture of the Attention-CNN model.}
\label{fig:attention_cnn}
\end{figure}

Following the idea from the previous section, we propose the Attention-CNN model, where the attention blocks are used as the electrode-sensitive component.
As shown in Figure \ref{fig:attention_cnn}, the Attention-CNN model is structured by adding an attention block after each convolution block in every layer and an additional single attention block before the final prediction block (the blocks in blue). A convolution block contains a convolution layer, a batch-norm layer \cite{ioffe2015batch}, a leaky ReLU \cite{maas2013rectifier} and a max-pooling layer. In addition, the residual \cite{he2016deep} techniques are applied in the CNN framework. The convolution layer operates only in the time dimension. The attention blocks, acting as an electrode-sensitive component, can be carried out by Squeeze-and-Excitation Block (SE Block) \cite{hu2018squeeze} and/or Self-Attention Block (SA Block) \cite{vaswani2017attention}. In the attention blocks, the retrieved electrode importance is used to weigh the features in each layer. Additionally, the same weights can provide explanations for the predictions of the model. In the prediction block, the features are flattened and then fed into the fully connected layer to finally obtain the predictions. 
While the SA Block is only required once in the process, the SE Blocks are added in every residual block. In order to keep the same scale for the same sample, the parameters of the SE Blocks are shared for the whole process. All building blocks are trained end-to-end, including the weights for the electrode importance used in the attention blocks.

\subsubsection{Squeeze and Excitation Block:}
the SE block involves two principle operations. The \textbf{Squeeze} operation compresses features $u \in \mathbb{R}^{T' \times J}$ into electrode-wise vectors $z \in \mathbb{R}^{J}$ by using global average pooling. Here, $T'$ denotes the feature size, and $J$ is the number of electrodes.  More precisely, the j-th element
of $z$ is calculated by $
  z_j = \mathbf{F}_{sq}(\mathbf{u}_j) = \frac{1}{T'} \sum_{i=1}^{T'} u_j(i)$.
The \textbf{Excitation} operation first computes activation $s$ by employing the gating mechanism with sigmoid activation: $s = \mathbf{F}_{ex}(\mathbf{z}, \mathbf{W}) = \sigma(\mathbf{W}_2 \delta(\mathbf{W}_1 \mathbf{z}))$, where $\sigma$ refers to the sigmoid function, $\delta$ represents the ReLU \cite{nair2010rectified} function, and $\mathbf{W}$ are learnable weights. The final output of SE block weigh each channel adaptively by re-scaling $U$ with $s$: $\tilde{\mathbf{x}_j} = \mathbf{F}_{scale}(\mathbf{u}_j, s_j) = s_j \cdot \mathbf{u}_j$. In contrast to the original implementation\cite{hu2018squeeze} which deals with 3-dimensional data, the input data in our setup has only 2 dimensions (electrodes and time).

\subsubsection{Self Attention Block:}
The \textit{self-attention} mechanism \cite{ribeiro2016should} was first used in the field of Natural language processing (NLP), aiming at catching the attention of/between different words in a sentence or paragraph. The attention is obtained by letting the input data interact with \textit{themselves} and determining which features are more important. This was implemented by introducing the \textit{Query, Key, Value} technique, which is defined as $\mathbf{Q} = \phi_Q(\mathbf{U},\mathbf{W}_Q)$, $\mathbf{K} = \phi_K(\mathbf{U},\mathbf{W}_K)$, $\mathbf{V} = \phi_V(\mathbf{U},\mathbf{W}_V)$, where $U$ denotes the input of self-attention block and $\phi(\cdot,\cdot)$ represents linear transformation. 

Then, \textit{Attention Weights} are computed using Query and Key:
$$\mathbf{M}_{att} = softmax(\frac{\mathbf{Q}\cdot \mathbf{K}^{T}}{\sqrt{d_k}})$$
where $d_k$ stands for the dimensions of the Key, and $\sqrt{d_k}$ works as a scaling factor. The softmax function was applied to adjust the range of the value in attention weights ($\mathbf{M_{att}}$) to $[0,1]$. 

Unlike the transformer model, the attention weights are first compressed into a one-dimensional vector by a layer of global average pooling ($\psi$) and normalized by a sigmoid function. More precisely, we compute $\mathbf{Z}_{att} = sigmoid(\psi(\mathbf{M}_{att}))$. Finally, the output of SA Block $\mathbf{X}$ is computed by 
: $\mathbf{X} = \kappa(\mathbf{Z}_{att},V)$, where $\kappa$ denotes the electrode-wise production.

\section{Experiments and Results}

\subsection{Materials and Experimental Settings}

\subsubsection{EEGEyeNet Dataset:}
For our experiments, we utilized the EEGEyeNet dataset \cite{kastrati2021eegeyenet}, which includes synchronized EEG and Eye-tracking data. 
The EEG signals were collected using a high-density, 128-channel EEG Geodesic Hydrocel system sampled at a frequency of 500 Hz. 
Eye-tracking data, including eye position and pupil size, were gathered using an infrared video-based eye tracker (EyeLink 1000 Plus, SR Research), also operating at a sampling rate of 500 Hz.
The recorded EEG and eye-tracking information was pre-processed, synchronized and segmented into 1-second clips
based on eye movements. The infrared eye tracking recordings were used as ground truth. In this paper, the processed dataset we utilized contains two parts: the \textit{Position Task} and \textit{Direction Task}, which correspond to two types of eye movements: \textit{fixation}, i.e.,  the maintaining of the gaze on a single location, and \textit{saccade}, i.e. the rapid eye movements that shift the centre of gaze from one point to another.
While \textit{Position Task} estimates the absolute position from fixation, \textit{Direction Task} estimates the relative changes during saccades, involving two sub-tasks, i.e., the prediction of amplitude and angle.
The statistics and primary labels of these two parts are shown in Table \ref{tab:dataset_description}.

\begin{table}
\caption{Dataset Description}\label{tab:dataset_description}
\begin{tabular}{c|c|c|l}
\hline
 \textbf{Task} &  \textbf{$\#$Subjects} & \textbf{$\#$Samples} & \textbf{Primary labels}\\
\hline
\multirow{2}{*}{Position} & \multirow{2}{*}{$72$} & \multirow{2}{*}{$41783$} & {\texttt{subject\_id}: the identical ID of the participant}\\
& & & \texttt{pos}: the fixation position in the form of $(x,y)$\\
\hline
\multirow{3}{*}{Direction} &  \multirow{3}{*}{72} & \multirow{3}{*}{50264} & {\texttt{subject\_id}: the identical ID of the participant}\\
& & & \texttt{amplitude}:the distance in pixels during the saccade\\
& & & \texttt{angle}: the saccade direction in radians\\

\hline
\end{tabular}
\end{table}

To ensure data integrity and prevent data leakage, the dataset was split into training, validation, and test sets across subjects, with 70 \% of the subjects used for training, and 15\%  each for validation and testing. This procedure ensures that no data from the same subject appears in both the training and validation/testing phases, thereby avoiding potential subject-related patterns from being learned by the model during training and tested on in validation/testing.
For more details of this dataset, please refer to \cite{kastrati2021eegeyenet}.

\subsubsection{Implementation Details:}
The experiments are implemented with PyTorch \cite{paszke2019pytorch}.
When training the Attention-CNN model, the batch size is set to 32, the number of epochs is 50, and the learning rate is $1e^{-4}$. There are 12 convolution blocks, and the residual operation repeats every three convolution blocks. The feature length of the hidden layer is set as 64, and the kernel size is 64. The number of convolutional layers, kernel size and hidden feature length, are selected based on validation performance.
We conducted experiments with three configurations: the SE Block and the SA Block together, only one of the attention blocks, or no attention blocks at all. For the angle prediction in Direction Task, we use angle loss $l_{angle} = |(atan(sin(p-t),cos(p-t))|$, where $p$ denotes the predicted results, and $t$ denotes the targets. For Position Task and Amplitude prediction in the Direction Task, the loss function is set to smooth-L1 \cite{girshick2015fast}.

\subsubsection{Evaluation:}
For Position task, Euclidean distance is applied as the evaluation metric in both pixels and visual angles. Compared to pixel distance, visual angles depend on both object size on the screen and the viewing distance, thus enabling the comparison across varied settings.
The performance of Direction Task is measured by the square root of the mean squared error (RMSE) for the angle (in radians) and the amplitude (in pixels) of saccades. In order to avoid the error caused by the repeatedness of angles in the plane (i.e. $2\pi$ and 0 radian represents the same direction), $atan(sin(\alpha),cos(\alpha))$ is applied, just like in angle loss. 


\subsection{Performance of the Attention-CNN}

Table \ref{tab:results} shows the quantitative performance of the Attention-CNN in this work. For the Position Task, CNN with SE block has an average performance with the RMSE of 109.58 pixels. Likewise, the CNN model with both SE block and the SA block has a similar performance (110.05 pixels).
Similar to Position Task, in amplitude prediction of Direction Task, the attention blocks aid the prediction evidently, heightening the performance by 5 pixels. Here, the model with both attention blocks has a lower variance. For angle prediction, the CNN model with both SE block and SA block has the best performance among all with the RMSE of 0.1707 radians.

We can conclude that the CNN model with both attention blocks consistently outperforms the CNN model alone by 5 to 10 percent across all tasks, indicating that electrode-wise attention assists in the learning process of the models.

\begin{table}[h]
  \begin{center}
    \caption{The performance of the Attention-CNN on Direction and Position Task.}
    \label{tab:results}
    \begin{tabular}{lcccc}
    \toprule
        \multirow{2}{*}{Models}& \multicolumn{2}{c}{\textit{Angle/Amplitude}} & & \textit{Abs. Position}\\\cmidrule{2-3}\cmidrule{5-5}
        & \textbf{Angle RMSE} & \textbf{Amp. RMSE} & & \textbf{Euclidean Distance (Visual Angle)}\\
        \midrule
        CNN & 0.1947±0.021 & 57.4486±2.053 && 115.0143±0.648 (2.39±0.010)\\
        CNN + SE & 0.1754±0.007 & 55.1656±3.513 & & \textbf{109.5816±0.238 (2.27±0.004)} \\
        CNN + SA  & 0.1786±0.010  & \textbf{52.1583±1.943}  & & 112.3823±0.851 (2.33±0.013)\\
        CNN + both & \textbf{0.1707±0.011} & \textbf{52.2782±1.169} & & \textbf{110.0523±0.670 (2.28±0.010)}\\
    \bottomrule
    \end{tabular}
  \end{center}
\end{table}

\subsection{Model Interpretability by Case Studies}

To provide a more detailed analysis of the interpretability of our proposed Attention-CNN model, as well as to further investigate the underlying reasons for the observed accuracy improvement, we conducted a visual analysis of the model performance, with a particular focus on the role of the attention block. Our analysis yielded two key findings, which are as follows:

\begin{figure}[t!]
  \centering
  \includegraphics[width=0.88\linewidth]{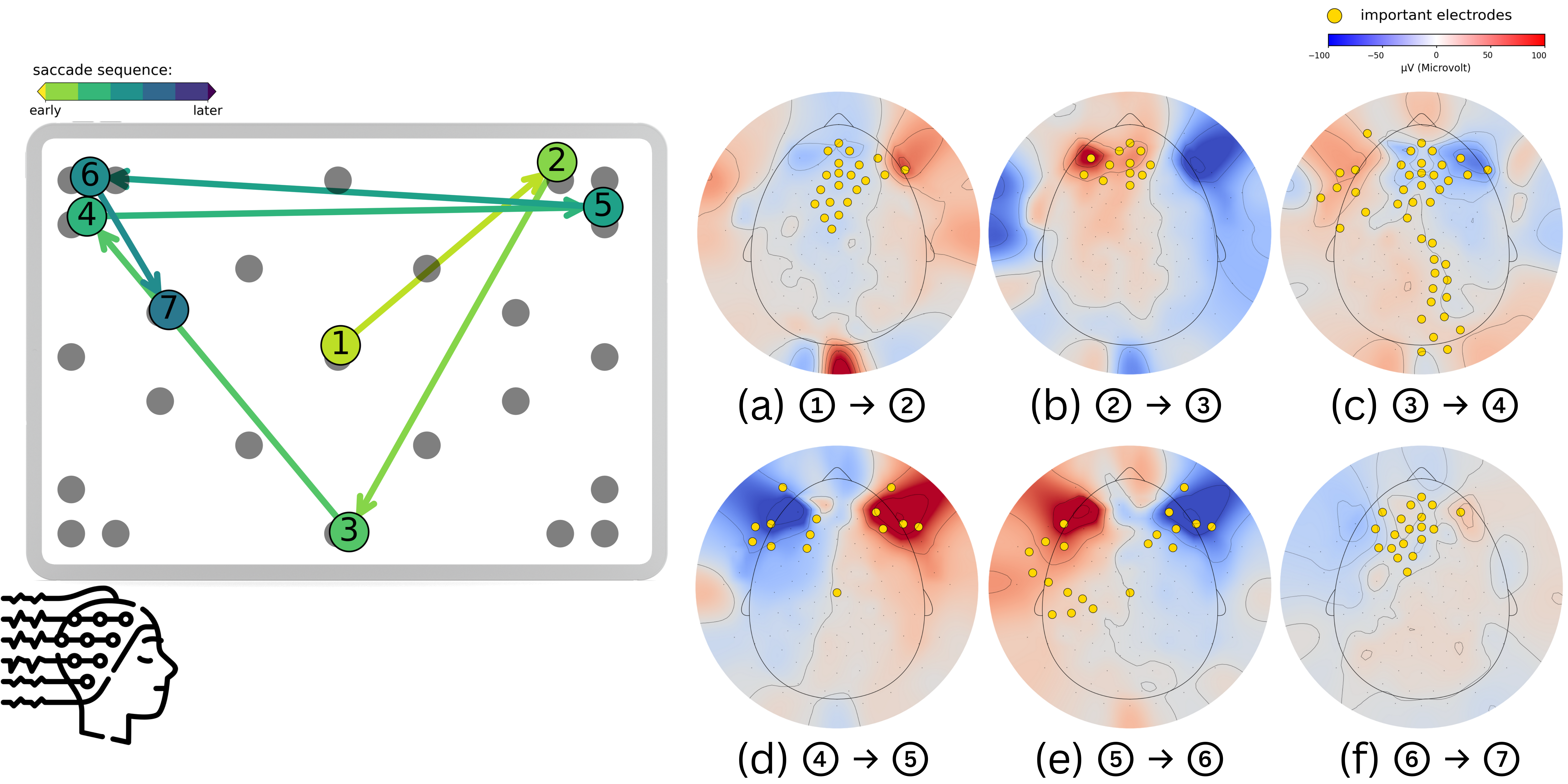}
  \caption{Visualization of signal intensity across scalp and electrode importance from our models. Left: the track of a  continuous sequence of saccades. Right: the corresponding brain activities (red: positive electrical signal, blue: negative electrical signal) and the important electrodes detected by the attention-based model (denoted as yellow nodes, the threshold is set as the mean value of all electrodes during the sequence). The model used here is the CNN with SA block.  }
  \label{fig:example_good}
\end{figure}

Firstly, the attention blocks were able to detect the electrical difference between the right and left pre-frontal area in case of longer saccades, i.e. rapid eye movements from one side of the screen to the other; see the saccades (d) and (e)
in Figure \ref{fig:example_good}. We present the sequence of saccades and observed the EEG signals as well as the electrode importance from proposed models in Figure \ref{fig:example_good}.
The attention block effectively captured this phenomenon by highlighting the electrodes surrounding the prominent signals (saccades (d) and (e) in Figure \ref{fig:example_good}). Conversely, in cases where the saccade was of a shorter distance (other saccades in Figure \ref{fig:example_good}), attention was more widely distributed across the scalp rather than being concentrated in specific regions. This is justifiable as the neural network aims to integrate a more comprehensive set of information from all EEG channels.

\begin{figure*}[t!]
    \centering
    \begin{subfigure}[t]{0.33\textwidth}
        \centering
        \includegraphics[height=1.4in]{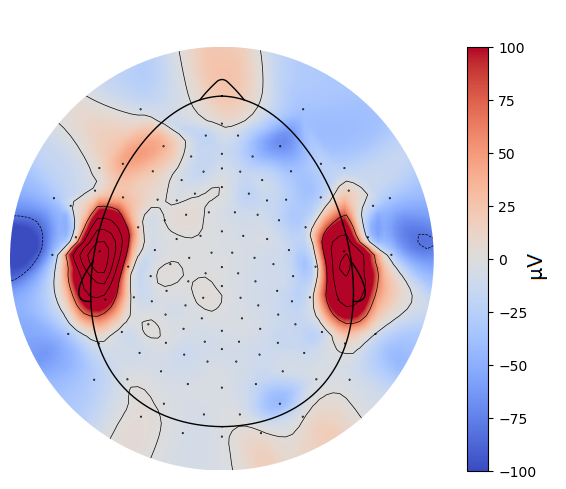}
        \caption{Input EEG data}
        \label{fig:example_problem-a}
    \end{subfigure}%
    \begin{subfigure}[t]{0.33\textwidth}
        \centering
        \includegraphics[height=1.4in]{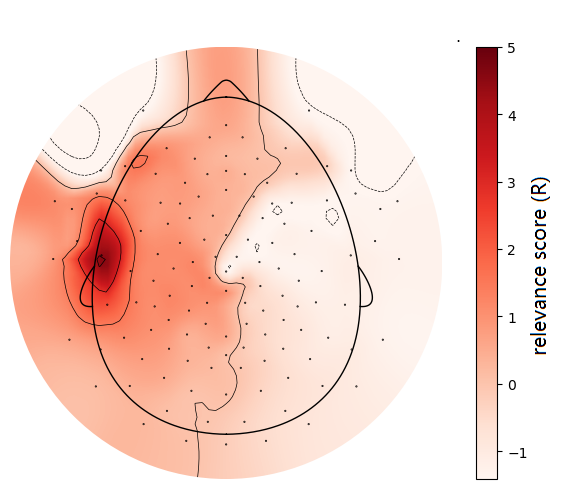}
        \caption{LRP results from CNN}
        \label{fig:example_problem-b}
    \end{subfigure}
    \begin{subfigure}[t]{0.33\textwidth}
        \centering
        \includegraphics[height=1.4in]{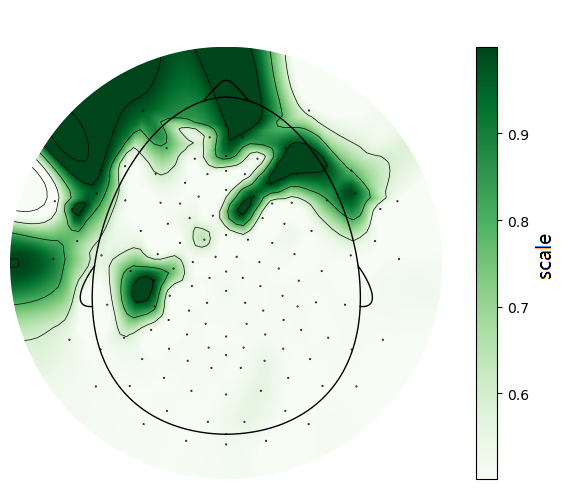}
        \caption{Scales from CNN+SA}
        \label{fig:example_problem-c}
    \end{subfigure}
    \caption{One example of test samples containing problematic electrodes is the Position Task. As shown in (a), the dark red areas around the ears represent intense electrical signals with abnormal amplitudes ($>100$ V). In (b), the Layer-wise Relevance Propagation (LRP) results from the CNN model reveal that the electrodes around the left ear still play a crucial role in the prediction process. Conversely, the Attention-CNN model's results (c), indicate that it bypasses the ear area and allocates more emphasis to the pre-frontal region. As a result, the error in Euclidean Distance improved by 200.85 pixels for this specific sample (from 265.18 to 64.33). }
    \label{fig:example_problem}
\end{figure*}

Additionally, the attention block effectively learned to circumvent the interference caused by noisy electrodes and redirected attention towards the frontal region. Figure \ref{fig:example_problem} illustrates a scenario where problematic electrodes were situated around both ears, exhibiting abnormal amplitudes ($\pm 100$ µV). Using Layer-wise Relevance Propagation\cite{bach2015pixel} to elucidate the CNN model's predictions, the result depicted in Figure \ref{fig:example_problem-b} revealed that the most significant electrodes were located over the left ear, coinciding with the noisy electrodes. In contrast, as shown in Figure \ref{fig:example_problem-c}, the Attention-CNN model effectively excluded the unreliable electrodes and allocated greater attention to the frontal region of the brain.

\subsection{Explainability Quantification}
We further examine the validity in explainability of the proposed method by comparing the distribution of learned attention of noisy and non-noisy electrodes in the Direction Task. The attention block's effectiveness is is demonstrated by its ability to assign lower weights to these noisy electrodes in contrast to the non-noisy ones.
Within all samples in the Direction Task that feature at least one noisy electrode, only 19\% of the non-noisy electrodes had normalized attention weights below $0.05$. In contrast, 42\% of the noisy electrodes exhibited this trait, implying the attention block's ability to reduce weights of abnormal electrodes. We direct readers to the Supplementary materials for a distribution plot showcasing the difference between noisy and non-noisy electrodes, along with additional details.
It's important to note that quantifying explainability methods for signal-format data, such as EEG, presents a significant challenge and has limited existing research. Therefore, additional investigations in this field are anticipated in future studies.

\section{Conclusion}

In this study, we aimed to address the issue of the lack of interpretability in deep learning models for EEG-based tasks. Our approach was to leverage the fact that EEG signal noise or artifacts are often localized to specific electrodes. We accomplished this by incorporating attention modules as electrode-sensitive components within a neural network architecture. These attention blocks were used to emphasize the importance of specific electrodes, resulting in more accurate predictions and improved interpretability through the use of scaling.

Moreover, our proposed approach was less susceptible to noise. We conducted comprehensive experiments to evaluate the performance of our proposed Attention-CNN model. Our results demonstrate that this model can accurately classify EEG and eye-tracking data while also providing insights into the quality of the recorded EEG signals. This contribution is significant as it can lead to the development of new decoding techniques that are less sensitive to noise.

In summary, our study underscores the importance of incorporating attention mechanisms into deep learning models for analyzing EEG and eye-tracking data. This approach opens up new avenues for future research in this area and has the potential to provide valuable insights into the neural basis of cognitive processes.

\bibliographystyle{splncs04}
\bibliography{mybibliography}

\begin{thebibliography}{10}
\providecommand{\url}[1]{\texttt{#1}}
\providecommand{\urlprefix}{URL }
\providecommand{\doi}[1]{https://doi.org/#1}

\bibitem{bach2015pixel}
Bach, S., Binder, A., Montavon, G., Klauschen, F., M{\"u}ller, K.R., Samek, W.:
  On pixel-wise explanations for non-linear classifier decisions by layer-wise
  relevance propagation. PloS one  \textbf{10}(7),  e0130140 (2015)

\bibitem{bulling2010eye}
Bulling, A., Ward, J.A., Gellersen, H., Tr{\"o}ster, G.: Eye movement analysis
  for activity recognition using electrooculography. IEEE transactions on
  pattern analysis and machine intelligence  \textbf{33}(4),  741--753 (2010)

\bibitem{carter2020best}
Carter, B.T., Luke, S.G.: Best practices in eye tracking research.
  International Journal of Psychophysiology  \textbf{155},  49--62 (2020)

\bibitem{craik2019deep}
Craik, A., He, Y., Contreras-Vidal, J.L.: Deep learning for
  electroencephalogram (eeg) classification tasks: a review. Journal of neural
  engineering  \textbf{16}(3),  031001 (2019)

\bibitem{devlin2018bert}
Devlin, J., Chang, M.W., Lee, K., Toutanova, K.: Bert: Pre-training of deep
  bidirectional transformers for language understanding. arXiv preprint
  arXiv:1810.04805  (2018)

\bibitem{duchowski2007eye}
Duchowski, A., Duchowski, A.: Eye tracking techniques. Eye tracking
  methodology: Theory and practice pp. 51--59 (2007)

\bibitem{eckstein2017beyond}
Eckstein, M.K., Guerra-Carrillo, B., Singley, A.T.M., Bunge, S.A.: Beyond eye
  gaze: What else can eyetracking reveal about cognition and cognitive
  development? Developmental cognitive neuroscience  \textbf{25},  69--91
  (2017)

\bibitem{feng2021automatic}
Feng, L.X., Li, X., Wang, H.Y., Zheng, W.Y., Zhang, Y.Q., Gao, D.R., Wang,
  M.Q.: Automatic sleep staging algorithm based on time attention mechanism.
  Frontiers in Human Neuroscience  \textbf{15},  692054 (2021)

\bibitem{girshick2015fast}
Girshick, R.: Fast r-cnn. In: Proceedings of the IEEE international conference
  on computer vision. pp. 1440--1448 (2015)

\bibitem{he2016deep}
He, K., Zhang, X., Ren, S., Sun, J.: Deep residual learning for image
  recognition. In: Proceedings of the IEEE conference on computer vision and
  pattern recognition. pp. 770--778 (2016)

\bibitem{holmqvist2012eye}
Holmqvist, K., Nystr{\"o}m, M., Mulvey, F.: Eye tracker data quality: What it
  is and how to measure it. In: Proceedings of the symposium on eye tracking
  research and applications. pp. 45--52 (2012)

\bibitem{hu2018squeeze}
Hu, J., Shen, L., Sun, G.: Squeeze-and-excitation networks. In: Proceedings of
  the IEEE conference on computer vision and pattern recognition. pp.
  7132--7141 (2018)

\bibitem{hu2022eeg}
Hu, Z., Chen, L., Luo, Y., Zhou, J.: Eeg-based emotion recognition using
  convolutional recurrent neural network with multi-head self-attention.
  Applied Sciences  \textbf{12}(21),  11255 (2022)

\bibitem{ioffe2015batch}
Ioffe, S., Szegedy, C.: Batch normalization: Accelerating deep network training
  by reducing internal covariate shift. In: International conference on machine
  learning. pp. 448--456. PMLR (2015)

\bibitem{kastrati2023electrode}
Kastrati, A., Plomecka, M.B., K{\"u}chler, J., Langer, N., Wattenhofer, R.:
  Electrode clustering and bandpass analysis of eeg data for gaze estimation.
  arXiv preprint arXiv:2302.12710  (2023)

\bibitem{kastrati2021eegeyenet}
Kastrati, A., P{\l}omecka, M.M.B., Pascual, D., Wolf, L., Gillioz, V.,
  Wattenhofer, R., Langer, N.: Eegeyenet: a simultaneous electroencephalography
  and eye-tracking dataset and benchmark for eye movement prediction. arXiv
  preprint arXiv:2111.05100  (2021)

\bibitem{lee2022eeg}
Lee, Y.E., Lee, S.H.: Eeg-transformer: Self-attention from transformer
  architecture for decoding eeg of imagined speech. In: 2022 10th International
  Winter Conference on Brain-Computer Interface (BCI). pp.~1--4. IEEE (2022)

\bibitem{maas2013rectifier}
Maas, A.L., Hannun, A.Y., Ng, A.Y., et~al.: Rectifier nonlinearities improve
  neural network acoustic models. In: Proc. icml. vol.~30, p.~3. Atlanta,
  Georgia, USA (2013)

\bibitem{munoz2004look}
Munoz, D.P., Everling, S.: Look away: the anti-saccade task and the voluntary
  control of eye movement. Nature Reviews Neuroscience  \textbf{5}(3),
  218--228 (2004)

\bibitem{nair2010rectified}
Nair, V., Hinton, G.E.: Rectified linear units improve restricted boltzmann
  machines. In: Icml (2010)

\bibitem{paszke2019pytorch}
Paszke, A., Gross, S., Massa, F., Lerer, A., Bradbury, J., Chanan, G., Killeen,
  T., Lin, Z., Gimelshein, N., Antiga, L., et~al.: Pytorch: An imperative
  style, high-performance deep learning library. Advances in neural information
  processing systems  \textbf{32} (2019)

\bibitem{ribeiro2016should}
Ribeiro, M.T., Singh, S., Guestrin, C.: " why should i trust you?" explaining
  the predictions of any classifier. In: Proceedings of the 22nd ACM SIGKDD
  international conference on knowledge discovery and data mining. pp.
  1135--1144 (2016)

\bibitem{ryan2010eye}
Ryan, J.D., Riggs, L., McQuiggan, D.A.: Eye movement monitoring of memory. JoVE
  (Journal of Visualized Experiments) (42),  e2108 (2010)

\bibitem{shaw2018self}
Shaw, P., Uszkoreit, J., Vaswani, A.: Self-attention with relative position
  representations. arXiv preprint arXiv:1803.02155  (2018)

\bibitem{sturm2016interpretable}
Sturm, I., Lapuschkin, S., Samek, W., M{\"u}ller, K.R.: Interpretable deep
  neural networks for single-trial eeg classification. Journal of neuroscience
  methods  \textbf{274},  141--145 (2016)

\bibitem{tabar2016novel}
Tabar, Y.R., Halici, U.: A novel deep learning approach for classification of
  eeg motor imagery signals. Journal of neural engineering  \textbf{14}(1),
  016003 (2016)

\bibitem{teplan2002fundamentals}
Teplan, M., et~al.: Fundamentals of eeg measurement. Measurement science review
   \textbf{2}(2),  1--11 (2002)

\bibitem{vachon2014eye}
Vachon, F., Tremblay, S.: What eye tracking can reveal about dynamic
  decision-making. Advances in cognitive engineering and neuroergonomics
  \textbf{11},  157--165 (2014)

\bibitem{vaswani2017attention}
Vaswani, A., Shazeer, N., Parmar, N., Uszkoreit, J., Jones, L., Gomez, A.N.,
  Kaiser, {\L}., Polosukhin, I.: Attention is all you need. Advances in neural
  information processing systems  \textbf{30} (2017)

\end{thebibliography}

\end{document}